\documentclass[12pt]{PoS}
\usepackage{graphics,graphicx,cite,color,amssymb,amsthm,amsmath}
\usepackage[update,prepend]{epstopdf} % to convert .eps files to .pdf
\usepackage{float}

\def\be{\begin{equation}}
\def\ee{\end{equation}}
\def\bea{\begin{eqnarray}}
\def\eea{\end{eqnarray}}

\def\bbuildrel#1_#2^#3{\mathrel{\mathop{\kern 0pt#1}\limits_{#2}^{#3}}}
\def\slash#1{\setbox0=\hbox{$#1$}#1\hskip-\wd0\dimen0=5pt\advance
       \dimen0 by-\ht0\advance\dimen0 by\dp0\lower0.5\dimen0\hbox
         to\wd0{\hss\sl/\/\hss}}

\newcommand{\f}{\frac}
\newcommand{\fm}[2]{{\textstyle \frac{#1}{#2}}}
\newcommand{\al}{\alpha_{\mathrm s}}
\newcommand{\alt}{\widetilde{\alpha}_{\mathrm s}}

\newcommand{\G}{\hat{G}}

     \title{On charm-mass dependent NNLO corrections to $\mathcal{B}(\bar{B}\rightarrow X_s \gamma)$}
\ShortTitle{On charm-mass dependent NNLO corrections to $\mathcal{B}(\bar{B}\rightarrow X_s \gamma)$}

\author{M. Misiak\\
        Institute of Theoretical Physics, Faculty of Physics,
        University of Warsaw, 02-093 Warsaw, Poland.\\
        E-mail: \email{mikolaj.misiak@fuw.edu.pl}}

\author{A. Rehman\speaker{}\\
        Institut f\"ur Theoretische Teilchenphysik, 
        Karlsruhe Institute of Technology (KIT), 76128 Karlsruhe, Germany.\\ 
        and \\
        National Centre for Physics, Quaid-i-Azam University Campus, Islamabad 45320, Pakistan.\\
        E-mail: \email{abdur.rehman@ncp.edu.pk}}

\author{M. Steinhauser\\
        Institut f\"ur Theoretische Teilchenphysik, 
        Karlsruhe Institute of Technology (KIT), 76128 Karlsruhe, Germany.\\
        E-mail: \email{matthias.steinhauser@kit.edu}}

      \abstract{The inclusive radiative decay of the $B$ meson is known to
        provide strong constraints on many popular extensions of the Standard
        Model. Such constraints crucially depend on precision of the Standard
        Model predictions. One of the main contributions to the theoretical
        uncertainty is due to certain Next-to-Next-to-Leading Order QCD
        corrections whose values at the physical charm quark mass $m_c$ have
        been estimated using interpolation between the $m_c=0$ and
        $m_c \gg m_b$ limits. A direct determination of such corrections at
        the physical value of $m_c$ requires calculating hundreds of
        two-scale four-loop propagator diagrams with unitarity cuts. Applying
        the integration-by-parts method, we express the corrections in terms
        of master integrals. Asymptotic expansions of these integrals at
        $m_c \gg m_b$ serve as boundary conditions for differential equations
        in $z=m_c^2/m_b^2$ that are being numerically solved. Here, we present
        our final results for the diagrams involving massless and massive
        fermion loops on the gluon lines. For the two-body cuts, we confirm
        the analytical expressions and/or numerical fits that are already
        present in the literature. In the four-body case, we make the
        correction complete by including several diagrams that have previously
        been only estimated using interpolation in $m_c$. We also report the
        status of the ongoing calculation of the remaining diagrams where no
        closed fermion loops on the gluon lines are present.}
\FullConference{RADCOR 2019 - 14th International Symposium on Radiative Corrections: Applications of Quantum Field Theory to Phenomenology\\ 
		9-13 September, 2019\\
		Avignon, France}

\begin{document}
\section{Introduction}

The flavour-changing neutral current transition $\bar{B}\rightarrow X_s \gamma$ proceeds via one-loop 
electroweak penguin diagrams at the leading order in the Standard Model (SM). It provides important constraints
on parameter spaces of many popular Beyond-SM (BSM) theories.
The current SM prediction for the CP- and isospin-averaged 
branching ratio with\footnote{ 
Such a conventional choice of the photon energy cut
is at the lower edge of the high-$E_0$ region where the experimental
background subtraction errors are manageable. At the same time, the
theoretical uncertainties are smaller than they would be for higher
$E_0$ where non-perturbative endpoint effects become significant.}
$E_\gamma > E_0 = 1.6\,$GeV has recently been updated~\cite{MRS2020}. It reads
${\mathcal B}_{s\gamma}^{\rm SM} = (3.40 \pm 0.17)\times 10^{-4}$, which should be compared\footnote{
The uncertainty reduction became possible thanks to a new
analysis of the so-called resolved photon contributions in
Ref.~\cite{Gunawardana:2019gep}, as well as to the improved isospin
asymmetry measurement by Belle~\cite{Watanuki:2018xxg}. These new
inputs are responsible for the central value shift, too.}
to the previous (2015) value of
${\mathcal B}_{s\gamma}^{\rm SM} = (3.36 \pm 0.23)\times 10^{-4}$ in Refs.~\cite{Misiak:2015xwa,Czakon:2015exa}.
Both values agree very well with the experimental world average
${\mathcal B}_{s\gamma}^{\rm exp} = (3.32 \pm 0.15)\times 10^{-4}$~\cite{Amhis:2019ckw}
that has been obtained from the results of CLEO~\cite{Chen:2001fja},
Babar~\cite{Aubert:2007my,Lees:2012wg,Lees:2012ym} and
Belle~\cite{Saito:2014das,Belle:2016ufb}, with extrapolation in $E_0$
down to $1.6\,$GeV. Constraints on BSM physics strongly depend on how
precisely the SM prediction is determined. For instance, with the
current precision, the resulting $95\%$ C.L. bound on the charged
Higgs boson mass in the Two-Higgs-Doublet Model~II calculated as in
Ref.~\cite{Misiak:2017bgg} is now in the vicinity of $800\,$GeV.

The present SM prediction uncertainty ($\pm 5\%$) has been 
determined by combining (in quadrature) uncertainties of three
different origins: $(i)$ parametric $(\pm 2.5\%$, including
non-perturbative effects), $(ii)$ higher-order perturbative $(\pm
3\%)$, and $(iii)$ interpolation in the charm quark mass that is used
to estimate some of the $\mathcal{O}(\al^2)$ corrections $(\pm
3\%)$~\cite{Czakon:2015exa}.  On the experimental side, the future
Belle II measurements are expected to eventually reduce the world
average uncertainty from the current $\pm 4.5\%$ to around $\pm
2.6\%$~\cite{Kou:2018nap,Ishikawa:2019Lyon}. Thus, the SM prediction
accuracy must be further improved to match the experimental
precision.

\begin{figure}[t]
\begin{tabular}{ccc}
\hspace{0.5cm} \includegraphics[scale=0.45]{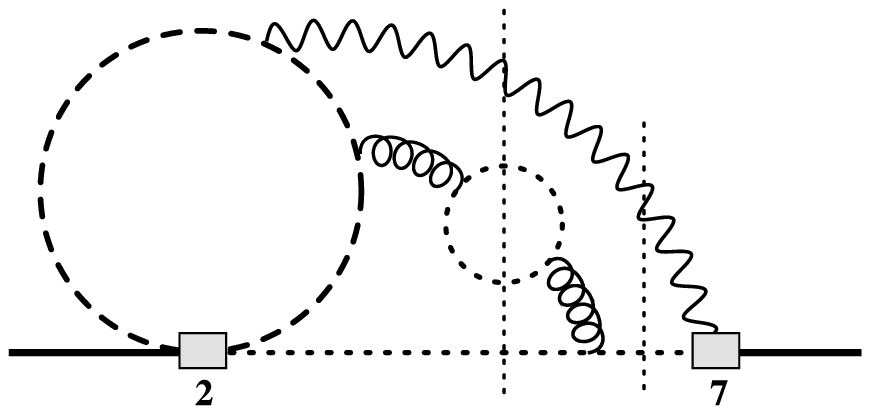}  \ \
&  \ \  \includegraphics[scale=0.45]{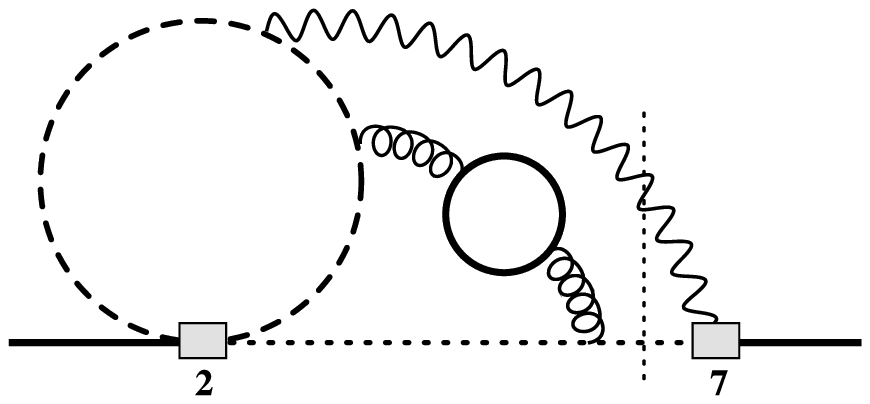} \ \
&  \ \  \includegraphics[scale=0.45]{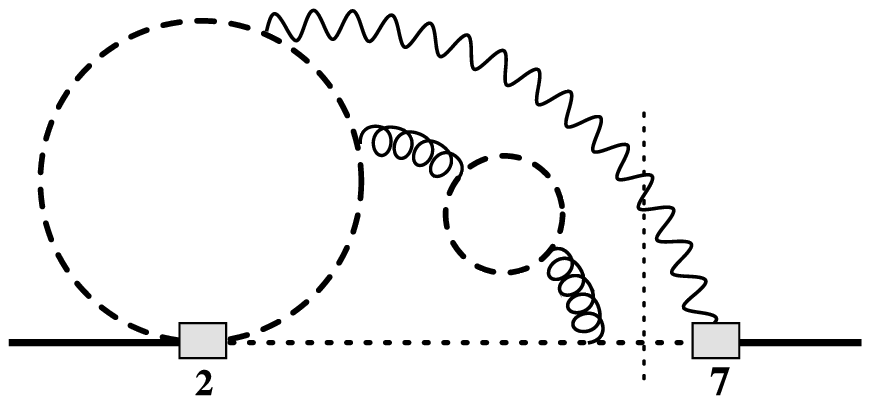}\\
\hspace{0.3cm}($\mathbf{a}$) [light quarks]&\hspace{0.3cm}($\mathbf{b}$) [bottom] &\hspace{0.3cm}($\mathbf{c}$) [charm]\\[5mm]
\end{tabular}
\begin{tabular}{ccc}
\hspace{0.5cm} \includegraphics[scale=0.45]{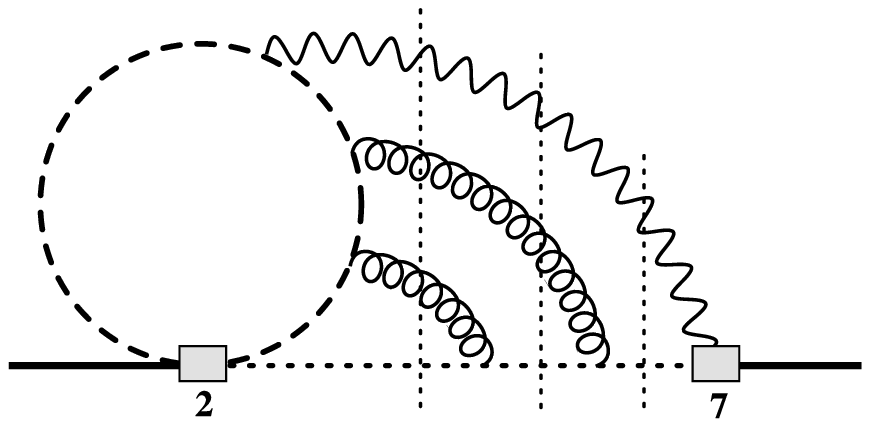}  \ \
&  \ \  \includegraphics[scale=0.45]{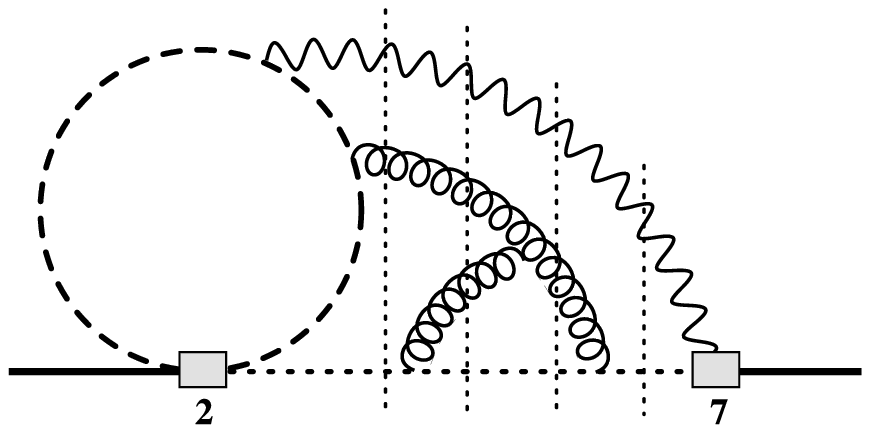} \ \
&  \ \  \includegraphics[scale=0.45]{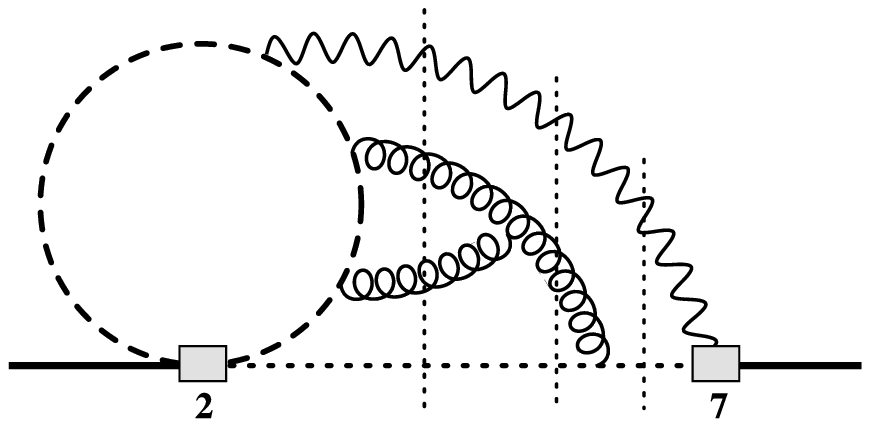}\end{tabular}
\caption{ \sf Sample type-I (first row) and type-II (second row) Feynman diagrams contributing to
$\G^{(2)}_{27}$. Possible unitarity cuts are indicated by the vertical
dotted lines.  The black solid, dashed and internal dotted lines
denote the $b$-quark, $c$-quark and $s$-quark propagators,
respectively. In the first diagram, all the light quarks ($u$, $d$ and $s$)
circulate in the closed loop on the gluon line.}  \label{fig:samplediagrams}
\end{figure}

Here, we focus on contributions that are necessary to remove 
the uncertainty related to the interpolation in $m_c$. To specify our object of interest, 
we begin with the perturbative rate of weak radiative $b$-quark decay
\be \label{def.ghat}
\Gamma(b \rightarrow X^{\rm partonic}_s \gamma) = 
\f{G_F^2 m^5_{b,pole}\alpha_{\mathrm em}}{32\pi^4} \left| V_{ts}^* V_{tb} \right|^2
\sum_{i,j} C_i(\mu_b) C_j(\mu_b) \G_{ij},
\ee
where $X^{\rm partonic}_{s}$ stands for $s, sg, sgg, sq\bar{q},
\ldots$ with $q=u,d,s$. For the decay rate evaluation, the Wilson
coefficients $C_i(\mu_b)$ that appear in the effective Lagrangian
${\mathcal L}_{\rm weak} \sim \sum_i C_i Q_i$ and the strong coupling
$\al$ are ${\overline{\rm MS}}$-renormalized at the low-energy scale
$\mu_b \sim m_b$. The quantities $\G_{ij}$ stand for interferences
of amplitudes generated by the effective operators $Q_i$ and
$Q_j$. They are perturbatively expanded as follows
\be
\G_{ij} = \G^{(0)}_{ij} + \alt \G^{(1)}_{ij} + \alt^2 \G^{(2)}_{ij} + {\mathcal O}(\al^3),
\ee
where $\alt = \al(\mu_b)/4\pi$. For our purpose, the following three operators $Q_i$ are relevant:
\be
Q_1  = (\bar{s}_L \gamma_{\mu} T^a c_L) (\bar{c}_L     \gamma^{\mu} T^a b_L),\hspace{8mm}
Q_2  = (\bar{s}_L \gamma_{\mu}     c_L) (\bar{c}_L     \gamma^{\mu}     b_L),\hspace{8mm}
Q_7  =  \fm{em_b}{16\pi^2} (\bar{s}_L \sigma^{\mu \nu}     b_R) F_{\mu \nu}.
\ee
We are interested in evaluating $\G^{(2)}_{17}$ and $\G^{(2)}_{27}$ at
the Next-to-Next-to-Leading-Order (NNLO) in QCD.  The considered
interferences can be represented in terms of propagator diagrams with
unitarity cuts corresponding to the two-, three- and four-particle
final states. These terms depend only on $E_0$, $\mu_b$, and the quark
mass ratio $z=m_c^2/m_b^2$, provided the light ($q=u,d,s$) quark
masses are neglected. Since the diagrams in $\G^{(2)}_{17}$ differ
from those in $\G^{(2)}_{27}$ by simple (diagram-dependent) colour factors
only, we shall discuss $\G^{(2)}_{27}$ alone in what follows. Sample
Feynman diagrams are shown in Fig.~\ref{fig:samplediagrams}. Altogether, around 350
%
% MM: 342
%
such diagrams need to be evaluated.

We express $\G^{(2)}_{27} (z, E_0)$ as
\be
\G^{(2)}_{27} (z, E_0) = \G^{(2), \text{type-I}}_{27} (z, E_0) + \G^{(2), \text{type-II}}_{27} (z, E_0), \label{type-InII}
\ee
where type-I parts of $\G^{(2)}_{27}$ arise from diagrams with closed
fermionic loops on the gluon lines, while the remaining contributions
are called type-II (see Fig.~\ref{fig:samplediagrams}). Analytical
and/or numerical results for $\G^{(2),\text{type-I}}_{27}(z,E_0)$ are
available from the calculations in Refs.~\cite{Ligeti:1999ea,Bieri:2003ue,Boughezal:2007ny,Misiak:2010tk},
except for a few diagrams\footnote{
Contributions from these diagrams were marked by $\kappa$ in
Ref.~\cite{Czakon:2015exa}, and estimated in the same way as type-II
ones.}
with four-body cuts presented in Fig.~3b of
Ref.~\cite{Czakon:2015exa}.  As far as type-II contributions are
concerned, the calculations have been so far finalized in two limiting
cases only. In Refs.~\cite{Misiak:2006ab,Misiak:2010sk},
$\G^{(2),\text{type-II}}_{27}(z,E_0)$ was determined for large $z$, at
the leading order in $1/z$. In Ref.~\cite{Czakon:2015exa},
$\G^{(2),\text{type-II}}_{27}(0,0)$ was calculated. Next, an
interpolation between these two limiting cases was performed to arrive
at an estimate for the considered correction at the physical value of
$m_c$, and with $E_0 = 1.6\,$GeV. The effect of the interpolated
$\mathcal{O}(\al^2)$ contribution on the branching ratio is shown in
Fig.~4 of Ref.~\cite{Czakon:2015exa}. As already mentioned, the
associated uncertainty has been estimated at the $\pm 3\%$ level,
which gives a significant contribution to the overall uncertainty of
the SM prediction. Therefore, evaluation of the considered correction
for the physical value of $m_c$ is important and necessary. Here, we
present our results for $\G^{(2),\text{type-I}}_{27}(z,0)$, and report
the status of the ongoing calculation of
$\G^{(2),\text{type-II}}_{27}(z,0)$.

Once $\G^{(2),\text{type-II}}_{27}(z,0)$ is found, the next step will
be to evaluate the difference between $\G^{(2)}_{27}(z,0)$ and
$\G^{(2)}_{27}(z,E_0)$ that comes from diagrams with three- and
four-body cuts only. However, for $E_0=1.6\,$GeV, this difference is
likely much smaller than $\G^{(2)}_{27}(z,0)$ itself, given that the
considered interference is peaked at the maximal $E_0$. At the NLO,
around $90\%$ of $\G^{(1)}_{27}(z,0)$ comes from the photons with
$E_\gamma > 1.6\,$GeV. Thus, the uncertainty stemming from the interpolation
in $m_c$ should essentially disappear after an explicit determination of
$\G^{(2),\text{type-II}}_{27}(z,0)$ alone.

\section{Evaluation of the master integrals for arbitrary ${\boldsymbol m_c}$}

After generating the Feynman diagrams with {\tt QGRAF}~\cite{Nogueira:1991ex} and/or
{\tt FeynArts}~\cite{Kublbeck:1990xc}, we perform the Dirac and colour algebra
with FORM~\cite{Ruijl:2017dtg} and self-written Mathematica codes,
respectively. At that stage, $G^{(2)}_{27}(z,0)$ is expressed in terms of
around $3\times 10^5$
%
% MM: 300758 (after removing the integrals that vanish under all the cuts)
%
four-loop two-scale scalar integrals in 437 families.  
Next, we use
{\tt KIRA}~\cite{Maierhofer:2018gpa} to perform the Integration-By-Parts
(IBP)~\cite{Tkachov:1981wb,Chetyrkin:1981qh,Laporta:2001dd} reduction.
In an alternative approach, we use {\tt QGRAF} together with {\tt q2e} and {\tt
  exp}~\cite{Harlander:1997zb,Seidensticker:1999bb} to generate a {\tt
FORM} code for the amplitudes, and perform the IBP reduction with 
 {\tt FIRE}~\cite{Smirnov:2019qkx} and {\tt LiteRed}~\cite{Lee:2013mka}.
For the most complicated families, several hundred~GB of RAM and weeks of
CPU are needed. Next, the Differential Equations
(DEs)~\cite{Kotikov:1990kg,Remiddi:1997ny,Gehrmann:1999as} 
\be \label{DEs}
\frac{\text{\bf d}}{\text{\bf d}z} {M}_n(z,\epsilon) 
= \sum_m R_{nm}(z,\epsilon) {M}_m(z, \epsilon)
\ee
are derived for the obtained Master Integrals (MIs)
$M_n(z,\epsilon)$. Getting a closed system of DEs often
requires including extra MIs. We numerically solve the DEs family-by-family,
without cross-mapping the MIs among different families. Within such an
approach, the total number of MIs is of order $10^4$. Such a large number of
MIs is not an obstacle, as our calculation is fully automatized.

The DE coefficients $R_{nm}(z,\epsilon)$ are rational functions of $z$
and the dimensional regularization parameter $\epsilon$. We expand in $\epsilon$,
and arrive at a DE system (analogous to the one in Eq.~(\ref{DEs})) for
functions of $z$ alone. Some of its coefficients usually contain poles on the
real axis. For this reason, our integration of the DEs proceeds along
ellipses in the complex $z$-plane, starting from initial conditions at
large $z$, similarly to the calculations in Refs.~\cite{Boughezal:2007ny,Misiak:2017woa,ARthesis}
\begin{figure}[t]
%\vspace*{-39mm}
\begin{center}
\includegraphics[width=11cm,angle=0]{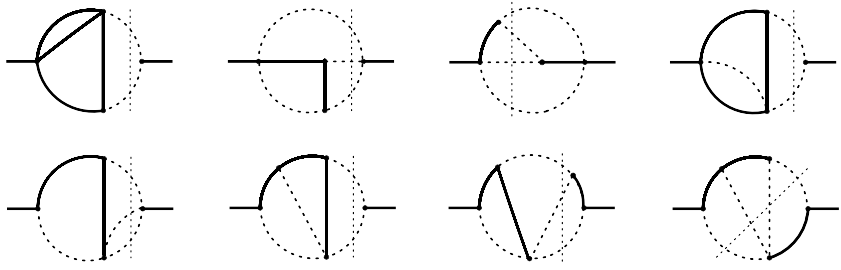}\\%[-4cm]
\caption{\sf Sample propagator-type integrals. The black solid and internal dotted lines denote the $b$-quark, 
and massless propagators, respectively. Unitarity cuts are shown by the thin dotted lines.
\label{fig:SSIs}}
\end{center}
\end{figure}

The initial conditions at large $z$ are evaluated using asymptotic expansions
for $m_c \gg m_b$. It leads to tadpole integrals up to three loops with
a mass scale $m_c$, as well as one-, two- and three-loop two-point integrals with
external momentum $q^2=m_b^2$, and internal lines that are either massless or
carry the mass $m_b$.  In the following, we denote such two-point integrals as
propagator-type integrals.  For the type-I contribution, we have to compute
two- and four-body cuts of the propagator-type integrals. In the 
type-II case, also three-body cuts are present. Sample integrals are shown in
Fig.~\ref{fig:SSIs}. The initial conditions are evaluated in an automatic
manner, using the code~{\tt exp}~\cite{Harlander:1997zb,Seidensticker:1999bb}.

\section{Results and progress \label{sec:results}}
One can write $\G^{(2), \text{type-I}}_{27} (z, 0)$ in Eq.~(\ref{type-InII}) as a 
sum of the following three contributions
\be
\G^{(2), \text{type-I}}_{27} (z, 0) = 3 \G^{(2), \text{type-I}(\bf a)}_{27} (z, 0) 
                                      + \G^{(2), \text{type-I}(\bf b)}_{27} (z, 0)  
                                      + \G^{(2), \text{type-I}(\bf c)}_{27} (z, 0),   \label{eq:type-Ia}
\ee
as depicted in the first row of Fig.~\ref{fig:samplediagrams}, for loops of the massless 
quarks ($u,d,s$), the bottom quark and the charm quark on the gluon propagator. 
After calculating the bare four-loop diagrams contributing to these quantities,
we renormalize them using the known counterterms~\cite{Misiak:2017woa,ARthesis}. 
One of the NLO contributions appearing in the counterterms is $\G^{(1)}_{47}$
stemming from the operator $Q_4 = (\bar{s}_L \gamma_{\mu} T^a b_L) \sum_q (\bar{q}\gamma^{\mu} T^a q)$.
Analytical expressions for this quantity presented in Eqs.~(2.4) and (B.1)-(B.2) of Ref.~\cite{Czakon:2015exa}
do not include charm-quark loops.\footnote{
The $z=0$ calculation in Section~2 of Ref.~\cite{Czakon:2015exa} focused on type-II contributions, as well as on
$\G^{(2),\text{type-I}(\bf a,b)}_{27}(0,0)$. However, $\G^{(2),\text{type-I}(\bf c)}_{27}(z,0)$ was included
in the arbitrary-$z$ expressions in Section 3 there.}
For the purpose of evaluating $\G^{(2),\text{type-I}(\bf
c)}_{27}(z,0)$, we have derived an analytical expression for the
charm-loop contribution to $\G^{(1)}_{47}$ up to the order ${\mathcal
O}(\epsilon)$. An explicit formula is presented in the appendix of 
Ref.~\cite{MRS2020}. 
\begin{figure}[t]
\begin{center}
\begin{tabular}{cc}
%\ & \ \\[-112mm]
\includegraphics[scale=0.52]{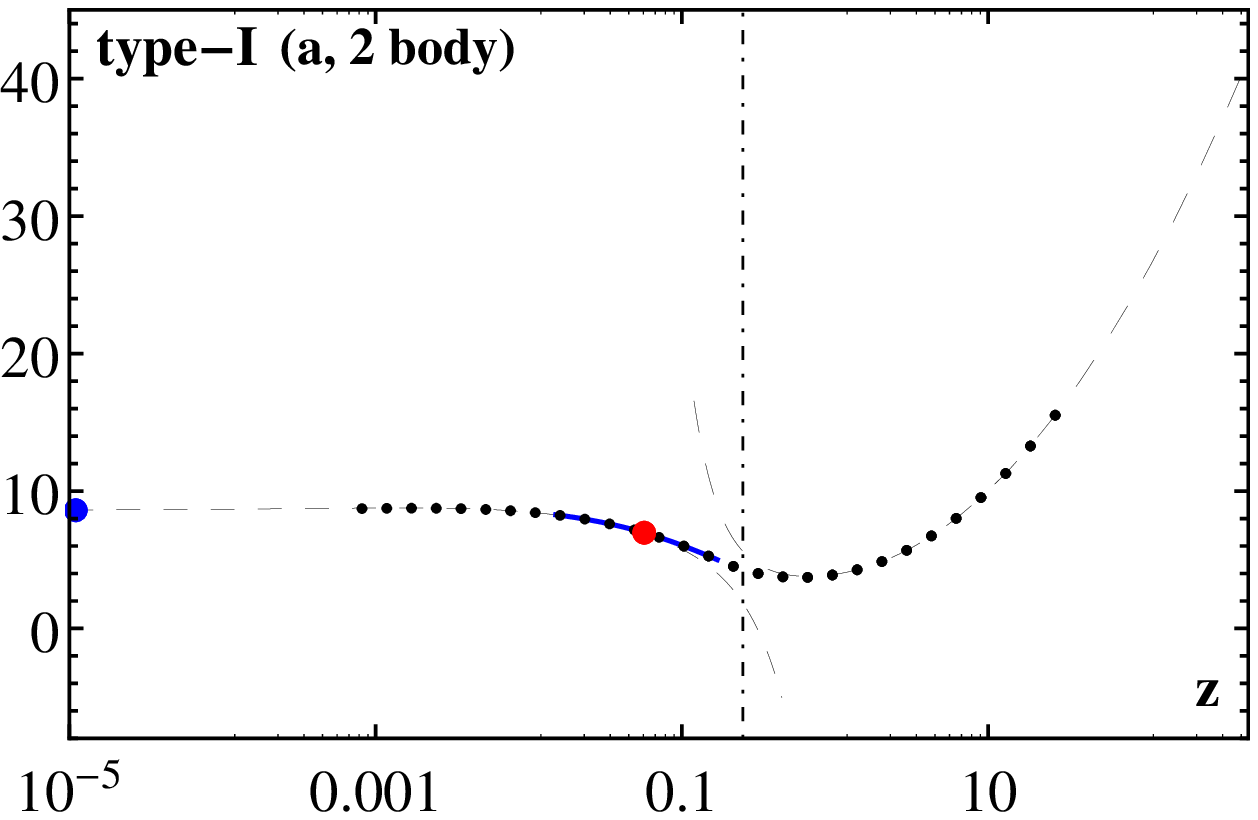} \hspace{0.3cm}
\includegraphics[scale=0.53]{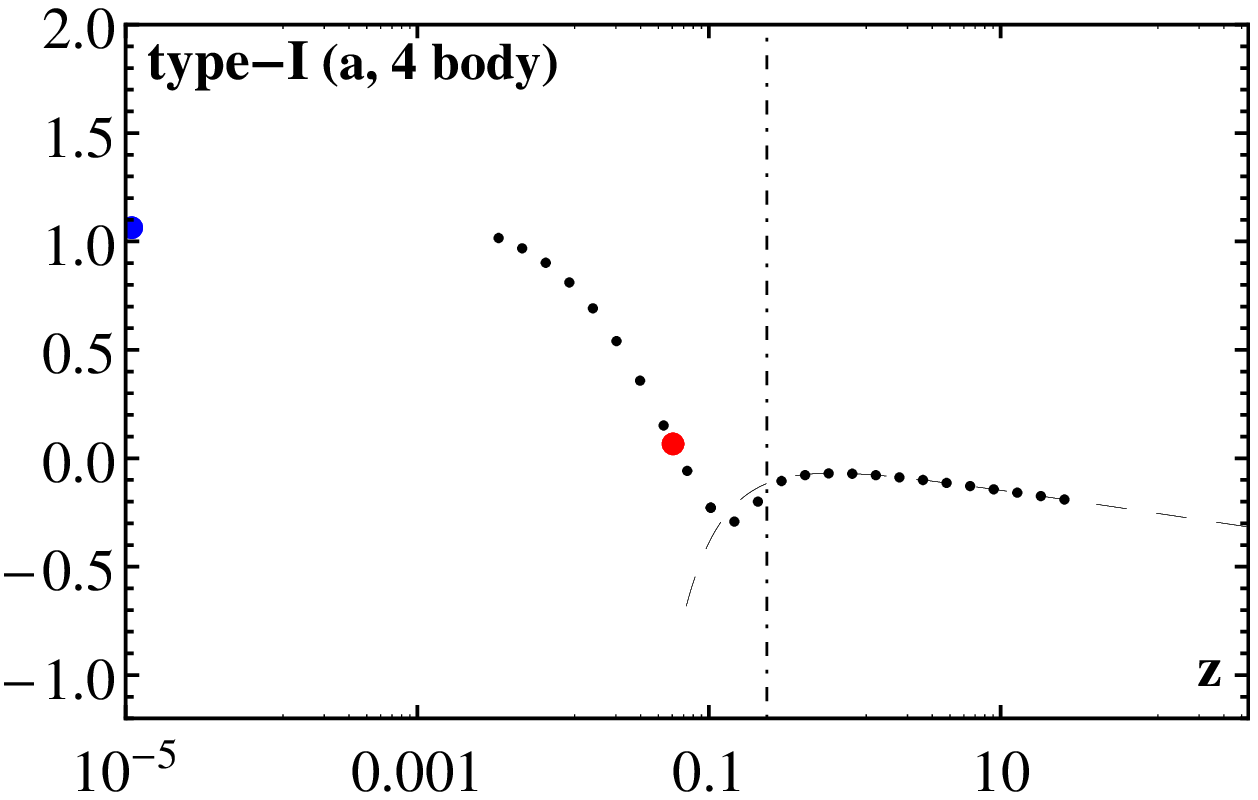}\\%[-107mm]
\includegraphics[scale=0.53]{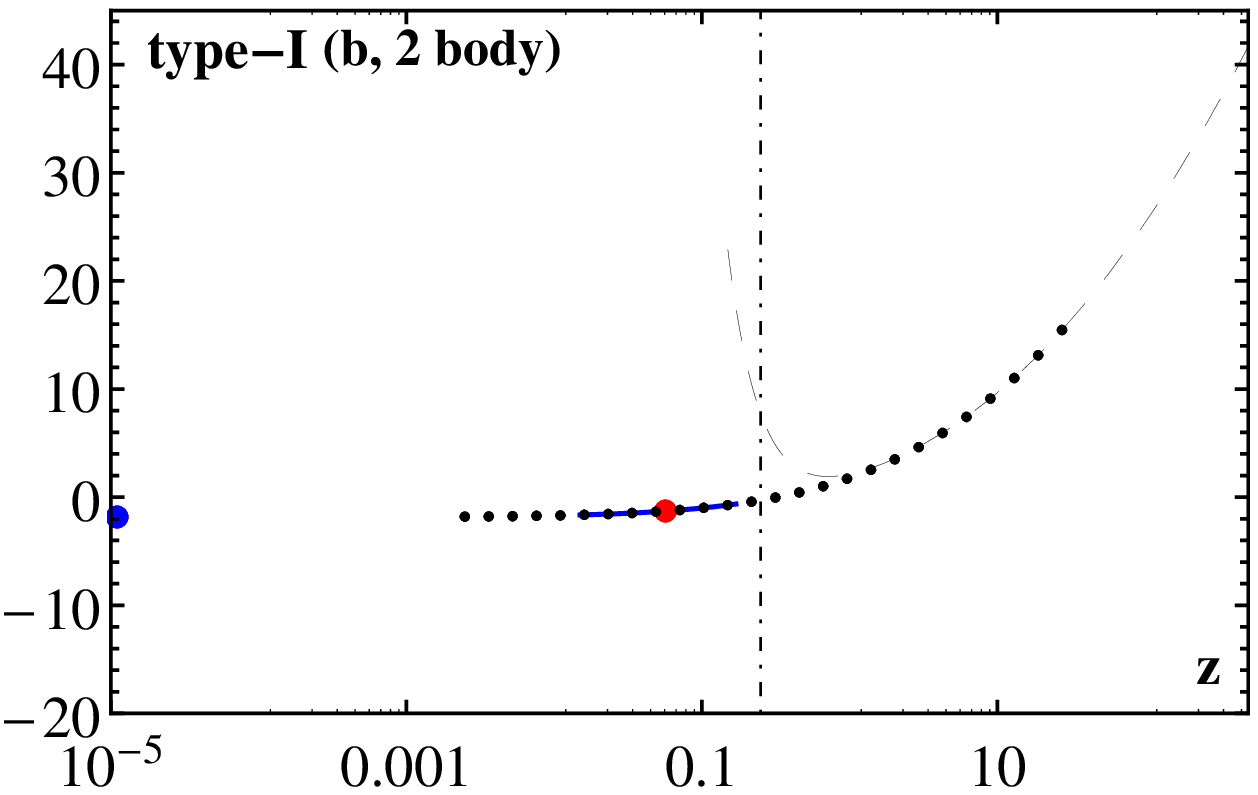} \hspace{0.3cm}
\includegraphics[scale=0.52]{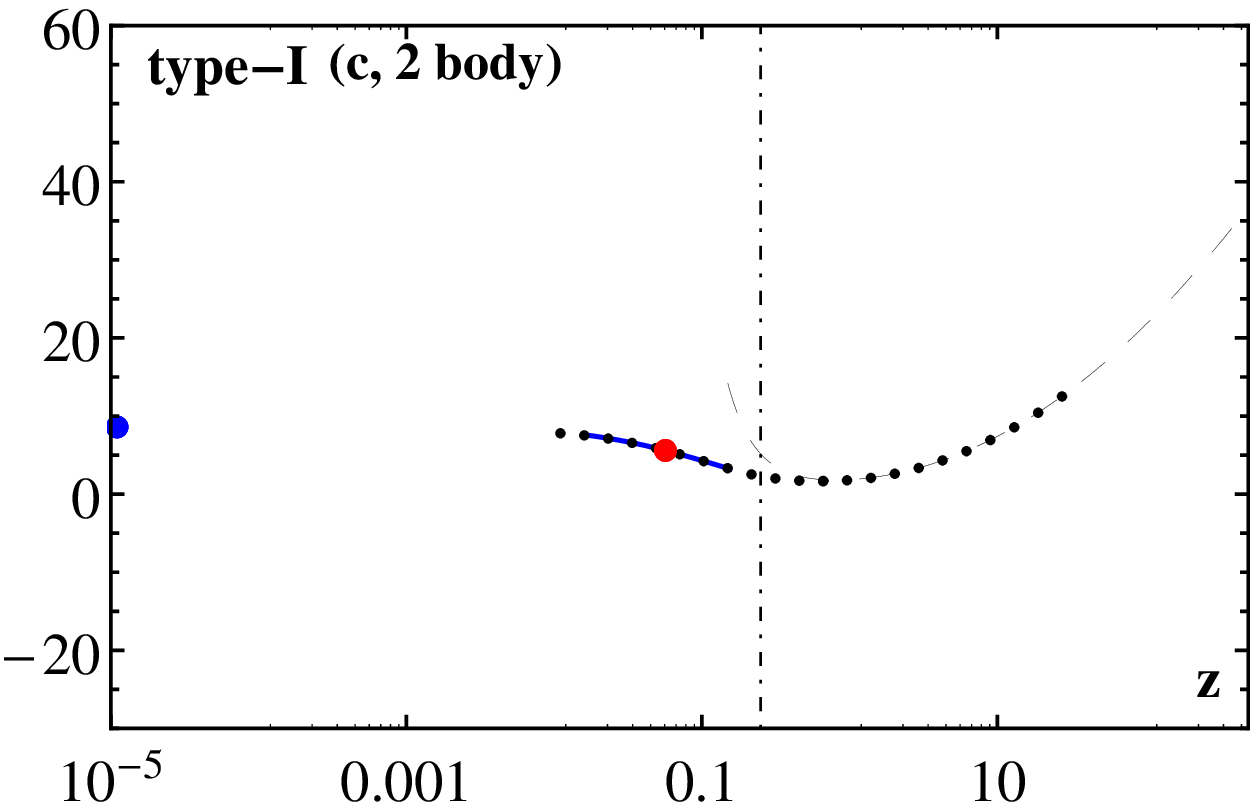}
\end{tabular}
\caption {\sf Plots of the functions defined in Eq.~(\ref{eq:type-Ia}). See the text for explanation.}
\label{fig:z-plots}
\end{center}
\end{figure}

Our ${\overline{\rm MS}}$-renormalized results for all the contributions to
$\G^{(2), \text{type-I}}_{27} (z, 0)$ in Eq.~(\ref{eq:type-Ia}) are displayed
in Fig.~\ref{fig:z-plots}, for the renormalization scale $\mu = m_b$. The
four-body part of $\G^{(2),\text{type-I}(\bf a)}_{27}(z,0)$ is understood to
contain also the three-body cut effects that enter via renormalization. As far
as $\G^{(2),\text{type-I}(\bf b,c)}_{27}(z,0)$ is concerned, only the two-body
parts are plotted -- the remaining parts, which are induced by
counterterm contributions, come in the term proportional to
$\phi^{(1)}_{27}(z,\delta)$ in Eq.~(3.8) of Ref.~\cite{Czakon:2015exa}.

In each plot of Fig.~\ref{fig:z-plots}, the black dots describe our
numerical solutions to the DEs. The dot corresponding to the
physical value of $z$ in the vicinity of $z=0.06$ is made larger and highlighted in
red. For each function, its limit at $z\to 0$ is indicated by a  blue
dot at the vertical axis on the left. These \linebreak
limits read~\cite{Czakon:2015exa}:~ 
$\left\{ {\bf a \text{ [2 body]}},\, {\bf a \text{[4 body]}},\, {\bf b \text{[2 body]}},\, {\bf c \text{[2 body]}} \right\}~ 
\stackrel{\scriptscriptstyle z\to 0\;\;}{\longrightarrow}~
\left\{ {N}_1,\, {N}_2,\, {N}_3,\, {N}_1 \right\}$,~ 
where \linebreak
${N}_1 = 2(33727-558\pi^2)/6561$,~ ${N}_2
\simeq 1.0640837328$,~ and ${N}_3 \simeq -1.8324081161$.  The
vertical dash-dotted lines indicate the $c\bar c$ production threshold at
$z=\f14$.  The dashed lines for $z > \f14$ are our large-$z$
expansions evaluated up to ${\mathcal O}(1/z^{2})$~\cite{MRS2020}.  Such expansions
with the MIs evaluated up to $\mathcal{O}(1/z^{5})$ have served as the
initial conditions at $z=20$ in our numerical solutions to the DEs.  The
dashed curve for $z < \f14$ in the upper-left plot is the known
small-$z$ expansion from Ref.~\cite{Bieri:2003ue}. The blue curves in
all the two-body plots are from the fit expressions published in
Ref.~\cite{Boughezal:2007ny}.

One can see that all our two-body results are in perfect agreement
with the previously published
ones~\cite{Bieri:2003ue,Boughezal:2007ny}.  In the four-body case, our
result is new, i.e. for the first time all the contributing diagrams have
been calculated for $z \neq 0$. The numerical solution in this case
gets very close to the $z=0$ limit when $z$ becomes as small as
$0.001$. For even lower $z$, numerical inaccuracies blow up, which we
can verify by testing cancellation of the coefficients at powers of
$1/\epsilon$ in the renormalization procedure. In
Fig.~\ref{fig:z-plots}, such a cancellation with a relative error of
better than $\mathcal{O}(10^{-3})$ has been required for all the
plotted black dots.  Our large-$z$ expansions and a numerical fit
for the new contribution are presented in Ref.~\cite{MRS2020}.

The convergence of our large-$z$ and small-$z$ expansions (dashed
lines) becomes poor in the vicinity of the threshold at $z=\f14$. It
is most visible in the plot of $\G^{(2), \text{type-I}(\bf
b)}_{27} (z,0)$. The same is true for the counterterm contributions
alone, in which case using the numerical results from
Refs.~\cite{Misiak:2017woa,ARthesis} (rather than the expansions)
allows us to successfully renormalize at points that are very close to
the threshold.

In the type-I case, we have a simple relation
$\G^{(2), \text{type-I}}_{17} (z) = -\f16 \G^{(2), \text{type-I}}_{27} (z)$,
which makes our plots in Fig.~\ref{fig:z-plots} directly applicable to the
$Q_1$-$Q_7$ interference, too.

As far as $\G^{(2), \text{type-II}}_{27} (z)$ is concerned, its evaluation is
in progress, following exactly the same lines as in the type-I case. The IBP
reduction and construction of the DEs have been completed. The boundary
conditions for $m_c\gg m_b$ are at the level of evaluation of three-loop
propagator-type integrals with two-, three- and four-particle cuts~\cite{SSI}.

\section{Summary}

We evaluated all the NNLO QCD corrections to
$\mathcal{B}(\bar{B}\rightarrow X_s \gamma)$ stemming from diagrams
with closed quark loops on the gluon lines, including cases where
the unitarity cut goes through such a loop. The calculation was
performed using the IBP method followed by numerically solving the DEs
for the MIs. Our results for the two-body final state contributions
are in agreement with the previous literature. In the four-body case,
our result includes contributions that have so far been only estimated
(using interpolation) at the physical value of $m_c$.

A calculation of the remaining (type-II) contributions for the physical $m_c$
is likely achievable using the same techniques as described in this work. It
is being carried out by a larger team~\cite{SSI}, currently focusing on evaluating
three-loop propagator-type integrals that parameterize the boundary conditions
for the DEs.

\section*{Acknowledgements} 

We would like to thank Johann Usovitsch and Alexander Smirnov for
their helpful advice concerning the use of {\tt KIRA} and {\tt
FIRE}, respectively. The research of AR and MS has been
supported by the Deutsche Forschungsgemeinschaft (DFG, German Research
Foundation) under grant 396021762 --- TRR 257 ``Particle Physics
Phenomenology after the Higgs Discovery''. MM has been partially
supported by the National Science Center, Poland, under the research
project 2017/25/B/ST2/00191, and the HARMONIA project under
contract UMO-2015/18/M/ST2/00518.

\end{document}